\documentclass[twocolumn]{autart}

\usepackage{float}
\usepackage{graphicx}
\usepackage{amsmath,amssymb}
\usepackage{subfigure}
\usepackage{dsfont}
\usepackage{psfrag}
\usepackage{algpseudocode}

\def\0{\mathbf{0}}
\def\1{\mathbf{1}}
\def\z{\mathbf{z}}

\def\I{{\mathbf I}}
\def\L{{\mathbf L}}
\def\v{{\mathbf v}}
\def\S{{\mathbf S}}
\def\p{{\mathbf p}}
\def\q{{\mathbf q}}

\def\V{{\mathbf V}}
\def\K{{\mathbf K}}
\def\J{{\mathbf J}}
\def\M{{\mathbf M}}
\def\W{{\mathbf W}}
\def\D{{\mathbf D}}

\def\B{{\mathbf B}}
\def\Q{{\mathbf Q}}
\def\R{{\mathbf R}}
\def\m{{\mathbf m}}

\def\b{{\bf b}}
\def\a{{\bf a}}
\def\u{{\bf u}}

\graphicspath{{figures/}}

\newtheorem{theorem}{Theorem}[section]

\newtheorem{proposition}[theorem]{Proposition}

\newtheorem{remark}[theorem]{Remark}

\floatstyle{ruled}
\newfloat{Algorithm}{tbp}{lop}[section]

\begin{document}
\begin{frontmatter}

\title{Hearing the Clusters of a Graph: A Distributed Algorithm}

\author[All]{Tuhin Sahai, Alberto Speranzon and Andrzej Banaszuk}\ead{SahaiT@utrc.utc.com, SperanA@utrc.utc.com and BansaszA@utrc.utc.com},

\address[All]{United Technologies Research Center, East Hartford, CT 06108, USA}

\begin{keyword}                           
Decentralized Systems, Graph Theory, Decomposition Methods              
\end{keyword}                             

\begin{abstract}
We propose a novel distributed algorithm to cluster graphs. The
algorithm recovers the solution obtained from spectral
clustering without the need for expensive eigenvalue/vector
computations. We prove that, by propagating waves through the
graph, a local fast Fourier transform yields the local component
of every eigenvector of the Laplacian matrix, thus providing
clustering information. For large graphs, the proposed algorithm
is orders of magnitude faster than random walk based approaches.
We prove the equivalence of the proposed algorithm to spectral
clustering and derive convergence rates. We demonstrate the
benefit of using this decentralized clustering algorithm for
community detection in social graphs, accelerating distributed
estimation in sensor networks and efficient computation of
distributed multi-agent search strategies.
\end{abstract}

\end{frontmatter}

\section{Introduction}

In recent years, there has been great interest in the analysis
of large interconnected systems, such as sensors networks,
social networks, the Internet, biochemical networks, power
networks, etc. These systems are characterized by complex
behavior that arises due to interacting subsystems. For such
systems graph theoretic methods have recently been applied and
extended to study these systems. In particular, spectral
properties of the Laplacian matrix associated with such graphs
provide useful information for the analysis and design of
interconnected systems. The computation of eigenvectors of the
graph Laplacian is the cornerstone of spectral graph
theory~\cite{Chung,Tutorial}, and it is well known that the
sign of the second (and successive eigenvectors) can be used to
cluster graphs~\cite{Fiedler,Fiedler2}.

The problem of graph (or data, in general) clustering arises
naturally in applications ranging from social
anthropology~\cite{Kottak}, gene networks~\cite{GeneSpectral},
protein sequences~\cite{ProteinSpectral}, sensor
networks~\cite{SensorNet,SensorNet2,Ali}, computer
graphics~\cite{CompGraph} and Internet routing
algorithms~\cite{KempeMcSherry08}.

The basic idea behind graph decomposition is to cluster nodes
 into groups with strong
intra-connections but weak inter-connections. If one poses the
clustering problem as a minimization of the inter-connection
strength (sum of edge weights between clusters), it can be
solved exactly and quickly~\cite{Stoer}. However, the
decomposition obtained is often unbalanced (some clusters are
large and others small)~\cite{Tutorial}. To avoid unbalanced
cuts, size restrictions are typically placed on the clusters,
i.e., instead of minimizing inter-connection strength, we
minimize the ratio of the inter-connection strength to the size
of individual clusters. This, however, makes the problem
NP-complete~\cite{Wagner}. Several heuristics to partition
graphs have been developed over the last few
decades~\cite{Communities} including the Kernighan-Lin
algorithm~\cite{Klin}, Potts method~\cite{Potts}, percolation
based methods~\cite{Percolation}, horizontal-vertical
decomposition~\cite{Igor} and spectral
clustering~\cite{Fiedler,Fiedler2}.

\subsection{Spectral clustering}

Spectral clustering has emerged as a powerful tool of choice for
graph decomposition purposes (see~\cite{Tutorial} and references
therein). The method assigns nodes to clusters based on the
signs of the elements of the eigenvectors of the Laplacian
corresponding to increasing
eigenvalues~\cite{Chung,Fiedler,Fiedler2}. In~\cite{Spielman},
the authors have developed a distributed algorithm for spectral
clustering of graphs. The algorithm involves performing random
walks, and at every step neglecting probabilities below a
threshold value. The nodes are then ordered by the ratio of
probabilities to node degree and grouped into clusters. Since
this algorithm is based on random walks, it suffers, in general,
from slow convergence rates.

Since the clustering assignment is computed using the
eigenvectors/eigenvalues of the Laplacian matrix, one can use
standard matrix algorithms for such
computation~\cite{GolubVanLoan96}. However, as the size of the
matrix (and thus the corresponding network) increases, the
execution of these standard algorithms becomes infeasible on
monolithic computing devices. To address this issue, algorithms
for distributed eigenvector computations have been
proposed~\cite{KempeMcSherry08}. These algorithms, however, are
also (like the algorithm in~\cite{Spielman}) based on the slow
process of performing random walks on graphs.

\subsection{Wave equation method}

In a theme similar to Mark Kac's question ``Can one hear the
shape of a drum?''~\cite{DrumShape}, we demonstrate that by
evolving the wave equation in the graph, nodes can ``hear'' the
eigenvectors of the graph Laplacian using only local
information. Moreover, we demonstrate, both theoretically and on
examples, that the wave equation based algorithm is orders of
magnitude faster than random walk based approaches for graphs
with large mixing times. The overall idea of the wave equation
based approach is to simulate, in a distributed fashion, the
propagation of a wave through the graph and capture the
frequencies at which the graph ``resonates''. In this paper, we
show that by using these frequencies one can compute the
eigenvectors of the Laplacian, thus clustering the graph. We
also derive conditions that the wave must satisfy in order to
cluster graphs using the proposed method.

The paper is organized as follows: in
Section~\ref{sec:related_work} we describe current methodologies
for distributed eigenvector/clustering computation based on the
heat equation. In Section~\ref{sec:wave_eqn} the new proposed
wave equation method is presented. In
Section~\ref{sec:perf_analysis} we determine bounds on the
convergence time of the wave equation. In
Section~\ref{sec:clustering_results} we show some numerical
clustering results for a few graphs, including a large social
network comprising of thousands of nodes and edges. We then
show, in Section~\ref{sec:estimation}, how the wave equation can
be used to accelerate distributed estimation in a large-scale
environment such as a building. In Section~\ref{sec:search} we
show how the proposed distributed clustering algorithm enables
one to efficiently transform a centralized search algorithm into
a decentralized one. Finally, conclusions are drawn in
Section~\ref{sec:conclusions}.

\section{From heat to wave equation: Related work}
\label{sec:related_work}

Let $\mathcal{G}=(V,E)$ be a graph with vertex set~$V =
\{1,\dots,N\}$ and edge set $E\subseteq V\times V$, where a
weight~$\W_{ij} \geq 0$ is associated with each edge $(i,j)\in
E$, and $\W$ is the $N\times N$ weighted adjacency matrix
of~$\mathcal{G}$. We assume that $\W_{ij}=0$ if and only if
$(i,j) \notin E$. The (normalized) graph Laplacian is defined
as,
\begin{align}
    \L_{ij} = \begin{cases}
                1 & \mbox{if}\: i = j\\
                -\W_{ij}/\sum_{\ell=1}^N \W_{i\ell} & \mbox{if}\: (i,j) \in E\\
                0   & \mbox{otherwise}\,,
             \end{cases}
             \label{eq:ldef}
\end{align}
or equivalently, $\L = \I-\D^{-1}\W$ where $\D$ is the
diagonal matrix with the row sums of $\W$.

Note that in this work we only consider undirected graphs. The
smallest eigenvalue of the Laplacian matrix is $\lambda_1 = 0$,
with an associated eigenvector
$\v^{(1)}=\1=\left[1,1,\dots,1\right]^T$. Eigenvalues of~$\L$
can be ordered as, $ 0 = \lambda_1 \leq \lambda_2 \leq \lambda_3
\leq \cdots \leq \lambda_N$ with associated eigenvectors $\1,
\v^{(2)}, \v^{(3)}\cdots \v^{(N)}$~\cite{Tutorial}. It is well
known that the multiplicity of~$\lambda_1$ is the number of
connected components in the graph~\cite{Mohar}. We assume in the
following that $\lambda_1<\lambda_2$ (the graph does not have
disconnected clusters). We also assume that there exist unique
cuts that divide the graph into~$k$ clusters. In other words, we
assume that there exist~$k$ distinct eigenvalues close to
zero~\cite{UlrikeDist}.

\begin{figure}[t!]
    \centering
    \includegraphics[width=0.85\hsize]{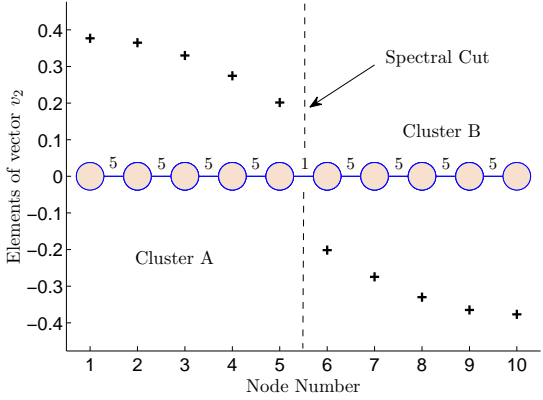}
    \caption{\label{ClusterEigen}Spectral clustering: The sign of the~$i$-th element of eigenvector~$v_2$ determines the cluster
    assignment of the~$i$-th vertex, demonstrated on a simple line graph example(shown in the center). With~$+$ we plot the value of the components of~$v_{2}$.}
\end{figure}

Given the Laplacian matrix~$\L$, associated with a
graph~$\mathcal{G}= (V,E)$, spectral clustering divides
$\mathcal{G}$ into two clusters by computing the sign of the~$N$
elements of the second eigenvector~$\v^{(2)}$, or Fiedler
vector~\cite{Tutorial,Fiedler2}. This process is depicted in
Fig.~\ref{ClusterEigen} for a line graph where one edge (the
edge (5,6)) has lower weight than other edges.

More than two clusters can be computed from signs of the
elements of higher eigenvectors, i.e.~$\v^{(3)}$, $\v^{(4)}$, etc.~\cite{Tutorial}. Alternatively, once the graph is
divided into two clusters, the spectral clustering algorithm can
be run independently on both clusters to compute further
clusters. This process is repeated until either a desired number
of clusters is found or no further clusters can be computed.
This method can also be used to compute a hierarchy of clusters.


There are many algorithms to compute eigenvectors, such as the
Lanczos method or orthogonal iteration~\cite{GolubVanLoan96}.
Although some of these methods are distributable, convergence is
slow~\cite{GolubVanLoan96} and the algorithms do not
consider/take advantage of the fact that the matrix for which
the eigenvalues and eigenvectors need to be computed is the
adjacency matrix of the underlying graph.
In~\cite{KempeMcSherry08}, the authors propose an algorithm to
compute the first~$k$ largest eigenvectors (associated with the
first~$k$ eigenvalues with greatest absolute value)\footnote{
Note that in the case of spectral clustering we desire to
compute the smallest $k$ eigenvectors of~$\L$. The algorithm is
still applicable if we consider the matrix~$\I-\L$.} of a
symmetric matrix. The algorithm in~\cite{KempeMcSherry08}
emulates the behavior of orthogonal iteration. To compute the
first~$k$ eigenvectors of a given matrix~$\J$, at each node in
the network, matrix~$\V_{i} = \sum_{j\in \mathcal{N}(i)} \J_{ij}
\Q_{j}$ is computed, where $\Q_{j}\in \mathds{R}^{N\times k}$ is
initialized to a random matrix and $\mathcal{N}(i)$ is the set
of neighbors of node~$i$ (including node~$i$ itself).
Orthonormalization is achieved by the computation of matrix
$\K_{i} = \V_{i}^{T}\V_{i}$ at every node, followed by
computation of matrix~$\K$, which is the sum of all the~$\K_{i}$
matrices in the network. Once matrix~$\K$ is computed, $\Q_{i} =
\V_{i}\R^{-1}$ is updated at each node, where~$\R$ is a unique
matrix such that~$\K = \R^{T}\R$ (Cholesky decomposition). The
above iteration is repeated until $\Q_{i}$ converges to
the~$i$-th eigenvector. The sum of all the matrices~$\K_{i}$ is
done in a decentralized way, using gossip~\cite{Shah09}, which
is a deterministic simulation of a random walk on the network.
In particular, at each node one computes the matrix~$\K$ as
follows,
\begin{align}
    \S_i(t+1) &= \sum_{j\in\mathcal{N}(i)}\B_{ji}\S_j(t)\,,\label{eq:gossip_matrix}\\
    \boldsymbol{\pi}_{i}(t+1) &= \sum_{j\in\mathcal{N}(i)}\B_{ji}\boldsymbol{\pi}_{j}(t)\,,\label{eq:gossip_mean}
\end{align}
for~$t \geq \tau$ steps, where~$\tau$ is the mixing time for the
random walk on the graph~\cite{KempeMcSherry08}. Here $\K =
\S_i/\boldsymbol{\pi}_{i}$, $\S_i(0) = \K_{i}$ and
$\boldsymbol{\pi}_{i}(0) = 1$ for only one index~$i$ and zero
for other indices. The values~$\B_{ij}$ are transition
probabilities of the Markov chain associated with the graph. A
natural choice is~$\B_{ij} = 1/\mbox{deg}(i)$, where
$\mbox{deg}(i)$ is the degree of node~$i$. Note that matrix $\B
= [\B_{ij}]$ is the normalized adjacency matrix (given by
$\D^{-1}\W$). This algorithm converges after $O(\tau\log^2 N)$
iterations~\cite{KempeMcSherry08}.

The slowest step in the distributed computation of eigenvectors
is the simulation of a random walk on the graph (defined by
Eq.~\ref{eq:gossip_matrix} and~\ref{eq:gossip_mean}). It is
clear from Eq.~\ref{eq:ldef} that successive multiplications by
the adjacency matrix $\B$ in Eqs.~\ref{eq:gossip_matrix}
and~\ref{eq:gossip_mean} are equivalent to successive
multiplications by matrix~$\I-\L$. This procedure is equivalent
to evolving the discretized heat equation on the graph and can
be demonstrated as follows. The heat equation is given by,
\begin{align*}
    \frac{\partial u}{\partial t} = \Delta u\,,
\end{align*}
where~$u$ is a function of time and space,~$\partial u/\partial
t$ is the partial derivative of~$u$ with respect to time,
and~$\Delta$ is the Laplace operator~\cite{Evans}. When the
above equation is discretized
(see~\cite{Chung,Hein,Hein2,BelNiyogi} for details) on a graph
$\mathcal{G}=(V,E)$ one gets the following equation:
\begin{align*}
    \u_{i}(t+1) = \u_{i}(t) - \displaystyle\sum_{j\in\mathcal{N}(i)}\L_{ij}\u_{j}(t)\,,
\end{align*}
for~$i,j \in V$. Here~$\u_{i}(t)$ is the scalar value of~$u$ on
node~$i$ at time~$t$. The graph Laplacian $\L=[\L_{ij}]$
appears due to the discretization of the $\Delta$
operator~\cite{Hein2}. The above iteration can be re-written,
in matrix form, $\u(t+1) = (\I-\L)\,\u(t)$ where~$\u(t) =
(\u_{1}(t),\dots,\u_{N}(t))^{T}$. The solution of this
iteration is,
\begin{align}
    \u(t) = C_0\1 + C_1 (1-\lambda_2)^t \v^{(2)} + \dots
    +C_N (1-\lambda_N)^t
    \v^{(N)}\,,
    \label{eq:HeatSol}
\end{align}
where constants~$C_{j}$ depend on the initial condition~$\u(0)$.
It is interesting to note that in Eq.~\ref{eq:HeatSol}, the
dependence of the solution on higher eigenvectors and
eigenvalues of the Laplacian decays with increasing iteration
count. Thus, it is difficult to devise a fast and distributed
method for clustering graphs based on the heat equation. Next,
we derive a novel algorithm based on the idea of permanent
excitation of the eigenvectors of~$\I-\L$. We note that the
above connection between spectral clustering and the heat
equation is not new and was pointed out in~\cite{Raphy1,Raphy2}.

Before discussing the details of wave-equation based eigenvector
computation, we remark that in~\cite{Seatzu} the authors have
independently developed a decentralized algorithm to compute the
eigenvalues of the Laplacian. Compared to our approach, their
algorithm involves solving a fourth order partial differential
equation on the graph. This imposes twice the cost of
communication, computation and memory on every node in the
graph.

\section{Wave equation based computation}
\label{sec:wave_eqn}

Consider the wave equation,
\begin{equation}
    \frac{\partial^{2} u}{\partial t^{2}}=c^{2}\Delta u\,.
    \label{Waveeqn}
\end{equation}
Analogous to the heat equation case (Eq.~\ref{eq:HeatSol}), the
solution of the wave equation can be expanded in terms of the
eigenvectors of the Laplacian. However, unlike the heat equation
where the solution eventually converges to the first eigenvector
of the Laplacian, in the wave equation all the eigenvectors
remain eternally excited~\cite{Evans} (a consequence of the
second derivative of $u$ with respect to time). Here we use this
observation to develop a simple, yet powerful, distributed
eigenvector computation algorithm. The algorithm involves
evolving the wave equation on the graph and then computing the
eigenvectors using local FFTs. Note that some properties of the
wave equation on graphs have been studied
in~\cite{WaveGraphProp}. Here we construct a graph
decomposition/partitioning algorithm based on the discretized
wave equation on the graph, given by
\begin{align}
    \u_{i}(t) = 2\u_{i}(t-1) - \u_{i}(t-2) - c^2\displaystyle\sum_{j\in\mathcal{N}(i)}\L_{ij}
    \u_{j}(t-1)\,,
    \label{onenodewave}
\end{align}
where $\sum_{j\in\mathcal{N}(i)}\L_{ij}\u_{j}(t-1)$ originates
from the discretization of $\Delta u$ in Eq.~\ref{Waveeqn},
see~\cite{Hein2} for details. The rest of the terms originate
from discretization of $\partial^{2} u/\partial t^{2}$. To
update~$\u_{i}$ using Eq.~\ref{onenodewave}, one needs only the
value of~$\u_{j}$ at neighboring nodes and the connecting edge
weights (along with previous values of~$\u_{i}$).

The main steps of the algorithm are shown as
Algorithm~\ref{alg:WaveAlg}. Note that at each node (node~$i$ in
the algorithm) one only needs nearest neighbor weights~$\L_{ij}$
and the scalar quantities $\u_{j}(t-1)$ also at nearest
neighbors. We emphasize, again, that $\u_{i}(t)$ is a scalar
quantity and \texttt{Random}($[0,1]$) is a random initial
condition on the interval $[0,1]$. The vector~$\v^{(j)}_{i}$ is
the $i$-th component of the $j$-th eigenvector, $T_{max}$ is a
positive integer derived in Section~\ref{sec:perf_analysis},
$\texttt{FrequencyPeak(Y,j)}$ returns the frequency at which the
$j$-th peak occurs and $\texttt{Coefficient}(\omega_{j})$ return the
corresponding Fourier coefficient.

\begin{Algorithm}[th!]
    \caption{Wave equation based eigenvector computation algorithm for node~$i$. At node~$i$ one computes the sign of the~$i$-th component of the first~$k$ eigenvectors. The cluster assignment is obtained by interpreting the vector of $k$ signs as a binary number.}\label{alg:WaveAlg}
    \begin{algorithmic}[1]
    \State $\u_{i}(0) \leftarrow \texttt{Random}\:([0,1])$
    \State $\u_{i}(-1) \leftarrow \u_{i}(0)$
    \State $t\leftarrow 1$
    \While{$t < T_{max}$}
         \State \begin{tabular}{l}
         $\u_{i}(t) \leftarrow 2\u_{i}(t-1)-\u_{i}(t-2) -$\\
         $\qquad \qquad c^2 \sum_{j\in\mathcal{N}(i)}\L_{ij}\u_{j}(t-1)$
         \end{tabular}
         \State $t\leftarrow t+1$
    \EndWhile
    \State $Y\leftarrow \texttt{FFT}\:(\left[\u_{i}(1),\dots\dots,\u_{i}(T_{max})\right])$
    \For{$j \in \{1,\dots,k\}$}
        \State $\omega_{j} \leftarrow \texttt{FrequencyPeak}\:(Y,j)$
        \State $\v^{(j)}_{i} \leftarrow \texttt{Coefficient}(\omega_{j})$ 
        \If {$\v^{(j)}_{i} > 0$}
            \State $A_j \leftarrow 1$
        \Else
            \State $A_j \leftarrow 0$
        \EndIf
        \EndFor
    \State ClusterNumber $\leftarrow \sum_{j=1}^k A_j 2^{j-1}$
\end{algorithmic}
\end{Algorithm}

\begin{proposition}
\label{prop:stability} The wave equation
iteration~\eqref{onenodewave} is stable on any graph if the wave
speed satisfies the following inequality,
$$
   0<  c < \sqrt{2}\,,
$$
with an initial condition of~$\u(-1) = \u(0)$.
\end{proposition}
\begin{pf}
For analysis of the algorithm, we consider Eq.~\ref{onenodewave}
in vector form,
\begin{align}
    \u(t) = -\u(t-2) + (2\I - c^{2}\L)\u(t-1)\,.
    \label{eq:wave_eq_iter}
\end{align}
We stress again that, in practice, the algorithm is distributed
and at every node one updates the state based on
Eq.~\ref{onenodewave}. The update equations given by
Eq.~\ref{onenodewave} (and Eq.~\ref{eq:wave_eq_iter}) correspond
to discretization of Eq.~\ref{Waveeqn} with Neumann boundary
conditions~\cite{Tomo1}.

One can write iteration Eq.~\ref{eq:wave_eq_iter} in matrix
form,
\begin{equation}
    \underbrace{\begin{pmatrix}
    \u(t) \\
    \u(t-1)
    \end{pmatrix}}_{\z(t)}=
    \underbrace{\begin{pmatrix}
    2\I - c^{2}\L & -\I \\
    \I & 0
    \end{pmatrix}}_\M
    \underbrace{\begin{pmatrix}
    \u(t-1)\\
    \u(t-2)
    \end{pmatrix}}_{\z(t-1)}\,.
    \label{uteqn}
\end{equation}
This implies that,
\begin{equation}
    \z(t) = \M^{t}\z(0)\,,
    \label{zoeq}
\end{equation}
where $\z(0) = (\u(0),\u(-1))^{T}$. We now analyze the solution
to Eq.~\ref{zoeq} in terms of the eigenvalues and eigenvectors
of the graph Laplacian~$\L$.

We can compute the eigenvectors of~$\M$ by solving for a generic
vector~$(\a_j,\b_j)^T$,
$$
    \M \begin{pmatrix}
        \a_j\\\b_j
        \end{pmatrix} = \alpha_j \begin{pmatrix}
        \a_j\\\b_j
        \end{pmatrix}\,.
$$
This implies that the eigenvectors of ~$\M$ are given by,
\begin{equation}
    \m^{(j)} =
    \begin{pmatrix}
        \alpha_j\: \v^{(j)}\\
        \v^{(j)}
    \end{pmatrix}\,,
    \label{eigenform}
\end{equation}
with eigenvalues
\begin{equation}
    \alpha_{j_{1,2}} = \frac{2-c^2\lambda_{j}}{2} \pm
    \frac{c}{2}\sqrt{c^2\lambda_{j}^2 - 4\lambda_{j}}\,.
    \label{alphaeqn}
\end{equation}
It is evident from Eq.~\ref{alphaeqn} that stability is obtained if
and only if,
$$
    \left| \frac{2-c^2\lambda_j}{2} \pm \frac{\sqrt{(2-c^2\lambda_j)^2-4}}{2}\right| \leq 1\,.
$$
The absolute value from the above equation is plotted for
various values of $\theta_j = 2-c^2\lambda_i$, in
Fig.~\ref{figabsval}. The above stability condition is satisfied
for $-2\leq \theta_j \leq 2$, which yields the following bound on~$c$:
$$
    0 \leq c \leq \frac{2}{\sqrt{\lambda_i}}\,.
$$
The above equation must hold true for all eigenvalues of $\L$.
The most restrictive of which is $c\leq 2/\sqrt{\lambda_{N}}$.
Since $\lambda_{N}\leq 2$ for all graphs,
\begin{equation}
    0 \leq c \leq \sqrt{2}\,,\nonumber
\end{equation}
guarantees that all the eigenvalues of $\M$ have absolute value
equal to one.
\begin{figure}[t!]
  \centering
  \psfrag{a}[][]{$\theta_j$}
  \psfrag{b}[b][]{$\left|\cfrac{\theta_j}{2} \pm \cfrac{\sqrt{(\theta_j)^2-4}}{2}\right|$}
  \includegraphics[width=0.85\hsize]{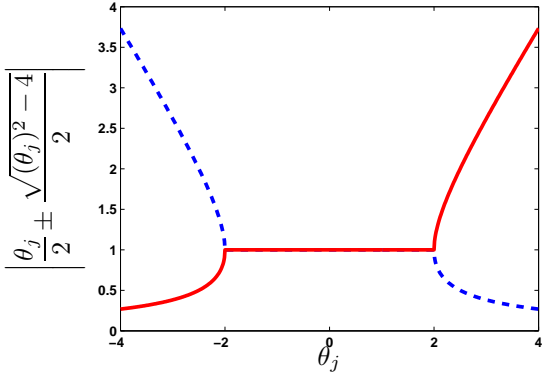}\\
  \caption{Plot of functions $|\theta_j/2 \pm 1/2 \sqrt{(\theta_j)^2-4}|$. Blue (dashed) line is the
function with a
  negative second term. Red (solid) line is the function with a positive second term.\label{figabsval}}
\end{figure}
However, Eqn.~\ref{onenodewave} will be unstable if any of the
eigenvalues of $\M$ have geometric multiplicity strictly less
than the algebraic multiplicity with an initial condition that
has non-zero projection on the unstable generalized
eigenvectors. We now derive conditions so that these
instabilities do not arise.

From Eqn.~\ref{alphaeqn} it is evident that there are three
cases to analyze.

\begin{description}
  \item[\emph{Case i)}:] Since~$\L$ always has an eigenvalue at~0, this implies that~$\M$ always
  has an eigenvalue at~1 with algebraic multiplicity two. It can be shown that the geometric multiplicity of this
  eigenvalue is equal to one.
  The corresponding eigenvector is $\1_{2N\times 1}$, with a generalized eigenvector $(\1,-\1)^{T}$. To avoid
  instability, the
  initial conditions must be of the form $(\u(0),\u(0))^{T}$. In other words, we
  set~$\u(-1)=\u(0)$ to ensure that the initial condition is
  orthogonal to $(\1,-\1)^{T}$.
  \item[\emph{Case ii)}:] If~$\L$ has~$k$ repeated eigenvalues, it implies that~$\M$ has~$k$ repeated eigenvalues.
   In this case, however, the geometric and algebraic multiplicities are equal. One can show that the matrix~$\L$ is
   similar to the symmetric matrix,
      $$
        \L_{sym} = \D^{-1/2}(\D-\W)\D^{-1/2}\,,
      $$
      (in particular, $\L = \D^{-1/2}\L_{sym}\D^{1/2}$), implying that $\L$ is diagonalizable. Thus, the
      eigenvectors of $\L$ associated with the
      repeated eigenvalues are linearly independent. Since matrix~$\M$ has eigenvectors of the form shown in
      Eqn.~\ref{eigenform}, the repeated eigenvalues of~$\M$ have eigenvectors that are linearly independent.
  \item[\emph{Case iii)}:] The matrix~$\M$ has a repeated eigenvalue at~-1 if ~$c^2 = 2$ and~$\lambda_N = 2$.
  This repeated eigenvalue has an associated eigenvector~$(-\v_N,\v_N)^T$
  and a
  generalized eigenvector~$(\v_N,\v_N)^T$. Clearly, in this case the initial condition would need to be
  orthogonal to both the vector~$(\1,-\1)^T$ and the vector~$(\v_N,\v_N)^T$. This can be achieved
  if and only if~$\u(0) \bot \v_N$ and $\u(-1)=\u(0)$. This is an undesirable condition, as it requires
  prior knowledge of~$\v_N$. We avoid this situation by setting $c<\sqrt{2}$.

\end{description}

Thus, we can guarantee stability of the wave equation iteration
on any graph (given by Eqn.~\ref{onenodewave}), as long as~$0< c
< \sqrt{2}$ and the initial condition has the form
$\u(-1)=\u(0)$.

Notice that the condition $\u(-1)\neq \u(0)$ is analogous to a
non-zero initial derivative condition on~$u$ for the continuous
PDE, which is known to give a solution that grows in
time~\cite{Evans}.
\end{pf}

\begin{remark}
Although we call~$c$ the wave speed, it only controls the extent
to which neighbors influence each other and not the speed of
information propagation in the graph.
\end{remark}

\begin{proposition}
The clusters of graph~$\mathcal{G}$, determined by the signs of
the elements of the eigenvectors of~$\L$, can be computed using
the frequencies and coefficients obtained from the Fast Fourier
Transform of $(\u_{i}(1),\dots,\u_{i}(T_{max}))$, for all~$i$
and some~$T_{max}>0$. Here~$\u_i$ is governed by the wave
equation on the graph with the initial condition~$\u(-1)=\u(0)$
and $0<c < \sqrt{2}$.
\end{proposition}

\begin{pf}
We can write the
eigenvectors~$\m^{(j)}$ of~$\M$ as,
\begin{equation}
    \m^{(j)} = \p^{(j)} \pm i \q^{(j)}\,, \nonumber
\end{equation}
where,
\begin{equation}
    \p^{(j)} = \begin{pmatrix}
    \mathrm{Real}(\alpha_{j})\v^{(j)}\\
    \v^{(j)}
    \end{pmatrix}\,, \quad \q^{(j)} = \begin{pmatrix}
    \textrm{Imag}(\alpha_{j})\v^{(j)}\\
    0
    \end{pmatrix}\,.\nonumber
\label{eq:pqvec}
\end{equation}
Using $\alpha_{j} = e^{i\omega_{j}}$, we can represent the
solution of the update equation (Eqn.~\ref{onenodewave}), or
equivalently,
\begin{equation}
\z(t) = \M^t\z(0)\,,
\label{eq:mteq}
\end{equation}
by expanding~$\z(0)$ in terms of $\p^{(j)}$ and $\q^{(j)}$.
Recall, that~$\z(0) = (\u(0),\u(0))^T$ is orthogonal to the
generalized eigenvector~$(\1,-\1)^T$. Thus,~$\z(0)$ is
represented as a linear combination of~$(\1,\1)^T$
and~$\m^{(j)}$ for $j\geq 2$. This implies that the solution to
Eqn.~\ref{uteqn} and~\ref{zoeq} is given by,
\begin{align}
    \z(t) = \sum_{j=1}^N
    C_{j_{1}}\left[\p^{(j)}\cos(t\omega_{j}) -
    \q^{(j)}\sin(t\omega_{j})\right] \nonumber \\
    +\;C_{j_{2}}\left[\p^{(j)}\sin(t\omega_{j}) +
    \q^{(j)}\cos(t\omega_{j})\right]\,, \label{Soln}
\end{align}
where
\begin{align}
    C_{j_{1}} = \z(0)^{T}\p^{(j)}\,,\quad C_{j_{2}} =
    \z(0)^{T}\q^{(j)}\,. \label{Ceqn}
\end{align}
It is easy to see that at every node, say the $i$-th node, one
can locally perform an FFT on
$(\u_{i}(1),\dots,\u_{i}(T_{max}))$ (where each value is
computed using the update law in Eq.~\ref{onenodewave}) to
obtain the eigenvectors. At the $i$-th node of the graph, one
computes the $i$-th component of every eigenvector from the
coefficients of the FFT. More precisely, for node~$i$, the
coefficient of $\cos(t\omega_{j})$ is given
$(C_{j_{1}}+C_{j_{2}})\v_{i}^{(j)}$. The sign of the
coefficients of the eigenvector(s) provide the cluster
assignment(s).

\end{pf}

%
%

\begin{remark}
The above algorithm assumes that one excites every frequency (or depending on the number of clusters, at least the
first~$k$ frequencies). This is achieved if~$\z(0)$ is not orthogonal to~$\p^{(j)}$ and~$\q^{(j)}$ ($C_{j_{1}}$
and $C_{j_{2}}$ must be non-zero).
As mentioned before, an initial condition of the form $\z(0) = (\u(0),\u(0))^{T}$ prevents linear growth of the solution,
however,~$\u(0)$ should also not be orthogonal to
$\v^{(2)},\v^{(3)}\dots \v^{(k)}$. This is easy to guarantee (with probability one) by picking a random initial
condition at each node.
\end{remark}

\begin{remark}
Note that the wave equation can also be used as a distributed
algorithm for eigenvector and eigenvalue computation of $\L$.
From the FFT we can compute~$\omega_{j}$ which in turn allows us
to
 compute the eigenvalues~$\lambda_{j}$. 
The eigenvector components are computed using the coefficients
of~$\cos(t\omega_j)$ (or equivalently $\sin(t\omega_{j})$).
\end{remark}

\begin{remark}
The algorithm is also attractive from a communication point of
view. In~\cite{KempeMcSherry08} entire matrices need to be
passed from one node to another. In our algorithm only scalar
quantities $\u_{j}$ need to be communicated.
\end{remark}

\begin{remark}
Peak detection algorithms based on the FFT are typically not
very robust because of spectral leakage. As we are only
interested in the frequencies corresponding to peaks, algorithms
like multiple signal classification~\cite{Music} can overcome
these difficulties. The investigation of such algorithms, as
well as windowing methods, is the subject of future work.
\end{remark}


\section{Performance analysis}
\label{sec:perf_analysis}

An important quantity related to the wave equation based
algorithm is the time needed to compute the eigenvalues and eigenvectors components. The distributed eigenvector algorithm proposed in~\cite{KempeMcSherry08} converges at a rate of $O(\tau\log^{2}(N))$ , where~$\tau$ is the
mixing time of the Markov chain associated with the random walk
on the graph. We derive a similar convergence bound for the
wave equation based algorithm.

It is evident from Eq.~\ref{Soln} that one needs to resolve the
lowest frequency to cluster the graph. Let us assume that one
needs to wait for~$\eta$ cycles of the lowest frequency to
resolve it successfully (i.e. the number of cycles needed for a
peak to appear in the FFT)\footnote{The constant~$\eta$ is
related to the FFT algorithm and independent of the graph.
Typically 6-7 cycles of the lowest frequency are necessary to
discriminate it.}. The time needed to cluster the graph based on
the wave equation is,
\begin{equation}
    T_{max} = \frac{\eta}{\omega_{2}}\,.
\end{equation}
From Eq.~\ref{alphaeqn} it is easy to see that $\cos(\omega_{2})
= \mathrm{Real}(\alpha_{2})= (2-c^2\lambda_{2})/2$. Note that
in~\cite{BoydMixing} it was shown that $\tau
=-(\log{|1-\lambda_{2}|})^{-1}$. Thus, it follows that,
$$
    \omega_{2} = \arccos\left(\frac{2+c^2
    (e^{-1/\tau}-1)}{2}\right)\,.
$$
Hence, the convergence of the wave equation based eigenvector
computation depends on the mixing time of the underlying Markov
chain on the graph, and is given by,
\begin{equation}
    T_{max} = O\left(\arccos\left(\cfrac{2+c^2 (e^{-1/\tau}-1)}{2}\right)^{-1} \right)\,.
    \label{eq:Waveconv}
\end{equation}
In the wave equation based clustering computation, one can at
the~$i$-th node, compute the~$i$-th component of every
eigenvector (along with all the eigenvalues) of the graph
Laplacian, thus assigning every node to a cluster.
If one uses the wave equation to compute eigenvectors, to ensure
that at every node one has entire eigenvectors, an extra
communication step needs to be added. As a final step, locally
computed eigenvectors components are transmitted to all other
nodes. The cost of this step is~$O(N)$ (worst case). Thus,
convergence of the distributed eigenvectors computation scales
as,
\begin{equation}
    T_{max} = O\left(\arccos\left(\cfrac{2+c^2 (e^{-1/\tau}-1)}{2}\right)^{-1} \right) +
    O(N)\,. \label{eq:Waveconvfinal}
\end{equation}
\begin{figure}[t!]
  \centering
  \includegraphics[width=0.85\hsize]{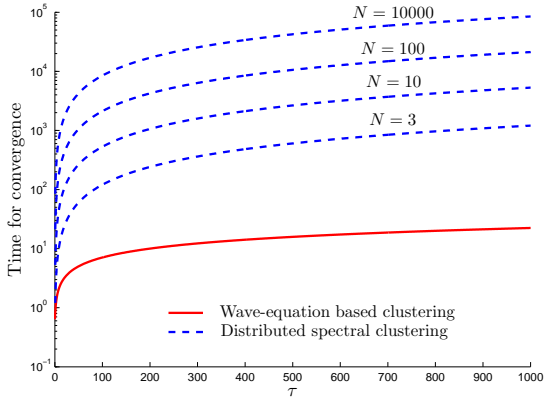}\\
  \caption{Comparison of convergence rates between the distributed algorithm in~\cite{KempeMcSherry08} and our proposed wave equation
  algorithm for~$c^{2}=1.99$. The wave equation based algorithm has better scaling with~$\tau$ for graphs of any size (given by $N$). The plots are upper bounds on the convergence speed.\label{ConvComp}}
\end{figure}
Note that simple analysis shows that for large~$\tau$ our
algorithm has a convergence rate of $\sqrt{\tau}/c$ (as $O(N)$
gets dominated by $\tau$). It is interesting to note that in the
discretized wave equation, though the constant~$c$ loses the
meaning of wave speed (that it has in the continuous case) it
does impact the speed of convergence.

The convergence of wave equation based clustering is compared to
convergence of distributed spectral clustering in
Fig.~\ref{ConvComp}, for~$c^{2} = 1.99$. In particular, the figure shows that wave equation based clustering has, in general,
better scaling, with
respect to~$\tau$, than~\cite{KempeMcSherry08}.

Note that the proximity of $\omega_{3}$ to $\omega_{2}$ (or the
proximity of $\lambda_{3}$ to $\lambda_{2}$) will influence the
constant in Eq.~\ref{eq:Waveconv}. The resolution of the FFT is
$O(1/K)$, where $K$ is the number of samples. Thus, $K$ has to
exceed $1/|\omega_{3}-\omega_{2}|$, to enable computation of two
separate peaks. The closer $\lambda_{3}$ is to $\lambda_{2}$,
the greater are the number of samples that each node needs to
store in order to obtain a good estimate of~$\omega_2$ using the
FFT. A similar constant depending on the ratio of $\lambda_{2}$
and $\lambda_{3}$ arises in distributed spectral
clustering~\cite{KempeMcSherry08} and any power iteration based
scheme for eigenvector computation~\cite{GolubVanLoan96}.

Practically, if the lowest frequency of the FFT does not change for a pre-defined length of time,
we assume that convergence has been achieved.

\begin{figure}[t!]
  \centering
  \includegraphics[width=0.55\hsize]{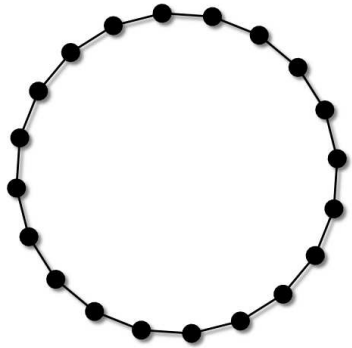}\\
  \caption{The ring graph $\mathcal{C}_N$ with~$N$ nodes. Every edge has a weight of~$1$.\label{Ringexample}}
\end{figure}

\begin{figure}[t!]
    \centering
    \includegraphics[width=0.85\hsize]{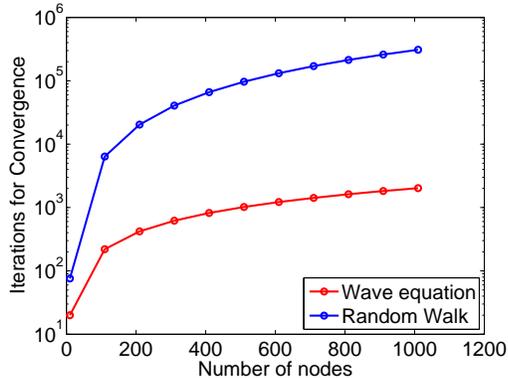}\\
    \caption{Convergence of random walk and wave equation on the cyclic graph~$\mathcal{C}_{N}$ as a function of number of nodes,~$N$.
    \label{ConvRing}}
\end{figure}

From Eq.~\ref{eq:Waveconv} it seems that the proposed clustering
algorithm is independent of the size of the graph (since
$\sqrt{\tau}/c$ dominates $O(N)$). This, however, is not true.
Larger graphs with low connectivity tend to have higher mixing
times. Take for example, a cyclic graph~$\mathcal{C}_N$ shown in
Fig.~\ref{Ringexample}. We use the cyclic graph as a benchmark
as one can explicitly compute the mixing time as a function of
$N$ and make a comparison with~\cite{KempeMcSherry08}. Of
course, no unique spectral cut exists for such a graph. The
second eigenvalue of the Laplacian for~$\mathcal{C}_N$ is given
by,
\begin{equation}
    \lambda_{2} = 1 - \cos\left(\frac{2\pi}{N}\right). \label{eq:Ringeig}
\end{equation}
Thus, the mixing time of the Markov chain is given by,
\begin{equation}
    \tau = -\frac{1}{\ln\left(\cos\left(2\pi/N\right)\right)}\approx \left(\frac{N}{2\pi}\right)^{2}\,.
    \label{eq:MixingTimeRing}
\end{equation}
From Eq.~\ref{eq:Waveconv}, one can show that the time for
convergence of the wave equation is,
\begin{align}
    T_{max} = \frac{\eta}{\arccos(1+0.5c^2(\cos(2\pi/N)-1))} \approx \eta\frac{N}{2\pi}\,. \label{eq:ConvRing}
\end{align}
As expected, Eq.~\ref{eq:ConvRing} predicts that as the graph
becomes larger, the convergence time for the wave equation based
algorithm increases. We numerically compute and compare the
convergence times for random walks and wave equation on the
cyclic graph (by explicitly running the iterations for both
processes and checking for convergence). The results are shown
in Fig.~\ref{ConvRing}.


\section{Numerical results}
\label{sec:clustering_results}

\begin{figure}[t!]
    \centering
    \includegraphics[width=0.85\hsize]{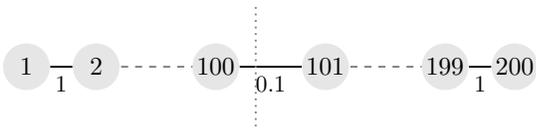}
    \caption{A line graph with nearest neighbor coupling.
    The edge between $100$ and $101$ is a weak connection with
    weight~$0.1$, all other edges have weight $1.0$. Vertical line
    shows the predicted cut.\label{Lineargraph}}
\end{figure}

\begin{figure}[t!]
    \begin{center}
        \includegraphics[width=0.85\hsize]{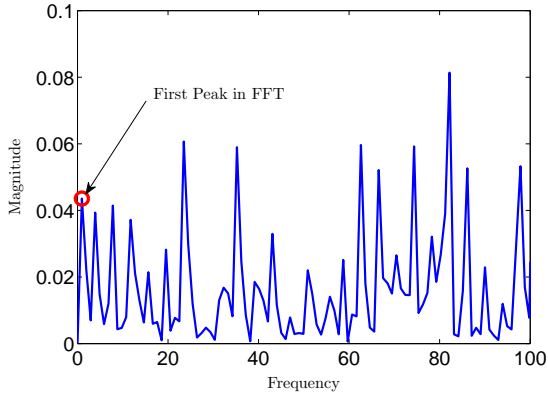}
    \end{center}
    \caption{\label{FFTlin} FFT of $\left[\u_{i}(1)\dots\u_{i}(T)\right]$ for any node $i$ of the line graph. Red circle marks the lowest frequency.}
\end{figure}

\begin{figure}[t!]
    \begin{center}
        \includegraphics[width=0.85\hsize]{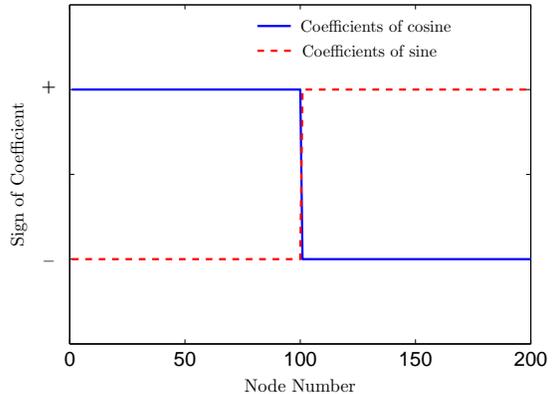}
    \end{center}
    \caption{\label{Coefflin} Signs of the coefficients of the lowest frequency for the line graph.}
\end{figure}

Since our algorithm should predict the same partitions as
spectral clustering, we demonstrate the algorithm on
illustrative examples. Our first example, is the simple line
graph shown in Fig.~\ref{Lineargraph}. Nodes $1$ to $100$ and
$101$ to $200$ are connected to their nearest neighbors with
edge weight $1$. The edge between nodes $100$ and $101$ has
weight $0.1$. As expected, spectral clustering predicts a cut
between nodes $100$ and $101$. We propagate the wave on the
graph using update Eq.~\ref{onenodewave} at every node. At each
node, one then performs an FFT on the local history of~$\u$. The
FFT frequencies are the same for all nodes (evident from
Eq.~\ref{Soln}) and shown in Fig.~\ref{FFTlin}. The sign of the
coefficients of the lowest frequency in the FFT are shown in
Fig.~\ref{Coefflin}. It is evident from this figure that the
sign of the coefficients change sign exactly at the location of
the weak connection, predicting a cut between nodes $100$ and
$101$ (consistent with spectral clustering).

We now demonstrate our distributed wave equation based
clustering algorithm on the Zachary Karate club
graph~\cite{Zachary} and on a Fortunato benchmark
example~\cite{Fortunato}. These social networks are defined by
the adjacency matrix that is determined by social interactions.
We assume that all the edges have weight $1$.

W. Zachary, a sociologist, was studying friendships at a Karate
club when it split into two. As expected, members picked the
club with more friends. This example serves as an ideal test bed
for clustering algorithms. Any effective clustering algorithm is
expected to predict the observed schism. Community detection and
graph clustering algorithms are routinely tested on this
example, see~\cite{Communities,NewZach,Newman,Rosvall} for a few
such demonstrations.

\begin{figure}[t!]
    \centering
\vspace{0.22in}
    \includegraphics[width=0.8\hsize]{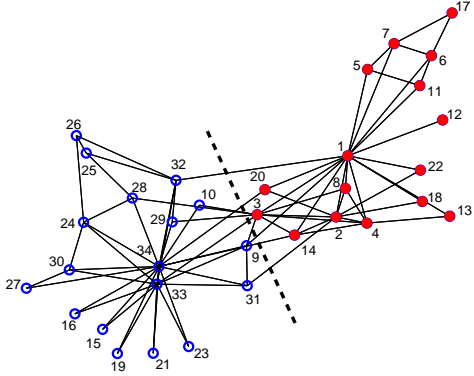}
    \caption{\label{karate}Graph decompositions predicted by spectral and wave equation based clustering algorithms. Both methods predict the same graph cut.}
\end{figure}

We first apply spectral clustering on this example, then run our
wave equation based clustering algorithm, and compare the
results in Fig.~\ref{karate}. As expected, we find that both
algorithms partition the graph into exactly the same clusters.

\begin{figure}[t!]
    \begin{center}
        \includegraphics[width=0.85\hsize]{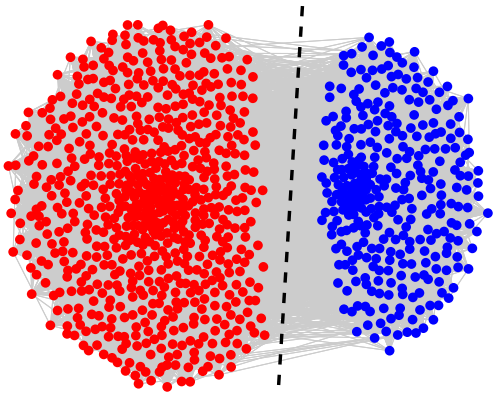}
    \end{center}
    \caption{\label{fig:Fortunato}A Fortunato community detection benchmark with $1000$ nodes and $99084$ edges. Wave equation based clustering computes the graph cut exactly.}
\end{figure}

We also demonstrate our algorithm on a large Fortunato benchmark
with $1000$ nodes and $99084$ edges. The graph has two natural
clusters with $680$ and $320$ nodes respectively. These clusters
are shown in Fig.~\ref{fig:Fortunato}. The wave equation based
clustering computes the graph cut exactly.

Thus, wave equation based eigenvector computation can be used to
partition both abstract graphs on parallel computers, or
physical networks such as swarms of unmanned vehicles, sensor
networks, embedded networks or the Internet. This clustering can
aid communication, routing, estimation and task allocation.

We now show how clustering can be effectively used to accelerate
distributed estimation and search algorithms.

\section{Distributed estimation over clusters}
\label{sec:estimation}

Distributed estimation has recently received significant
attention
see~\cite{Alrikson2007}--\nocite{Olfati-Saber2007,Carli2008b}\cite{Speranzon2008}
and references therein. Distributed estimation algorithms
require the entire network of sensors to exchange (through
nearest neighbor communication) data about the measured
variables in order to obtain an overall estimate, which is
asymptotically (in the number of iterations) optimal. This
results in estimators with error dynamics that converge to zero
very slowly. It is well known that these type of algorithm can
be accelerated using multi-scale approaches, see for
example~\cite{JHK08,JHK08-CDC,SelleWest09}. The key idea in
these multi-scale approaches is to partition the sensor network
into clusters, solve the distributed problem in each cluster and
fuse the information between clusters.

As the overall estimation process is distributed, it is
desirable that the multi-scale speedup is achieved through a
distributed process as well. This means that the clustering must
be computed, in a bottom-up fashion, from the structure of the
network. We show in the following a simple yet illustrative
example, where the wave equation based clustering algorithm can
be used to accelerate distributed estimation computation by
exploiting properties of the overall sensor network.

\begin{figure}[t!]
    \centering
    \includegraphics[width=0.95\hsize]{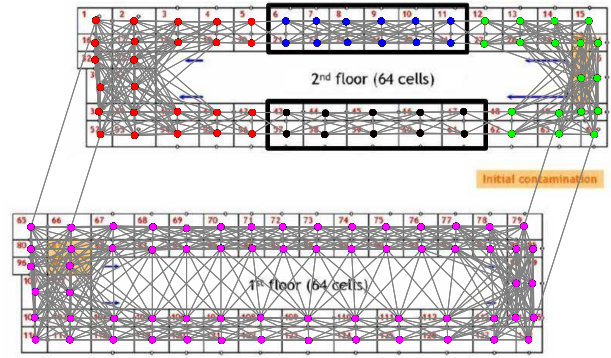}\\
    \caption{A two floor building subdivided into 64 cells/rooms for each floor. In each room there is a sensor node capable of communicating with neighbors within a radius of 10m. The thick black line, depicts walls that degrade communication strength.}\label{fig:building}
\end{figure}

We consider the contaminant transport problem in a
building~\cite{JHK08} with two floors, each divided into 64
cells/rooms (see Fig.~\ref{fig:building}). A sensor, to detect
the contaminant, is present in each cell. Sensors can
communicate if their relative distance is less than 10m.
However, we assume that only four sensors can communicate
between floors, namely those placed within common staircases
connecting the two floors. On the first floor, sensors can
communicate across the empty space in between (we assume that
windows are present), whereas on the second floor we assume that
there are walls that reduce the communication range. We further
assume that walls marked with a thick black line, see
Fig.~\ref{fig:building}, degrade communication between the nodes
that are inside the area to those outside.
As in~\cite{JHK08}, we assume that the contaminant is produced
in four rooms, two on the first floor and two on the second.
Under the simplifying assumption of perfect mixing within each
cell/room volume, the contaminant propagates within the building
(see~\cite{JHK08} for details) according to:
\begin{align*}
    \rho_iV_i&\frac{dC_i}{dt}=\sum_{i\sim j}{F_{ji}C_j}-
    \sum_{i\sim j}{F_{ij}C_i}+G_i-R_iC_i\,,\nonumber\\
    \rho\text{ : }&\text{Density}\qquad  C\text{ : }\text{Contaminant concentration}\\
    V\text{ : }&\text{Volume} \qquad  F_{ji}\text{ : }\text{Mass flow rate from node $j$ to $i$}\nonumber\\
    G\text{ : }&\text{Contaminant generation rate}\nonumber\\
    R\text{ : }&\text{Mass removal rate}\,.
\end{align*}
A constant inward flow of air is introduced at a corner of the
second floor, and outflow openings exist wherever windows are
open to the outside. We consider a distributed Kalman filter as
one that uses consensus to average the estimates and covariance
matrices between Kalman filter updates, see~\cite{JHK08} for
details.

The idea of using the wave equation for distributed clustering
is to ``discover'' in a bottom-up fashion, the presence of
clusters and exploit strong inter-cluster connectivity to
accelerate computation. In particular, we demonstrate the
benefit of using the ``bottom-up'' approach. In the building
example there are two main clusters (first and second floors),
which a filter and network designer can a-priori assume to know.
The four clusters on the second floor, however, would not be
known to the designer unless extensive communication
measurements are carried out.

In order to determine the four clusters based on SNR, on the
second floor, the wave equation based clustering was run for 600
steps. The clustering clearly needs to be run only once, unless
there is very strong variation of SNR or the network. In this
particular example we assume that the SNR and the network do not
vary.

\subsection{Numerical results}
\label{sec:estimation_results} 
\begin{figure*}[t!]
    \centering
    \subfigure[Two Clusters]{\includegraphics[width=0.45\hsize]{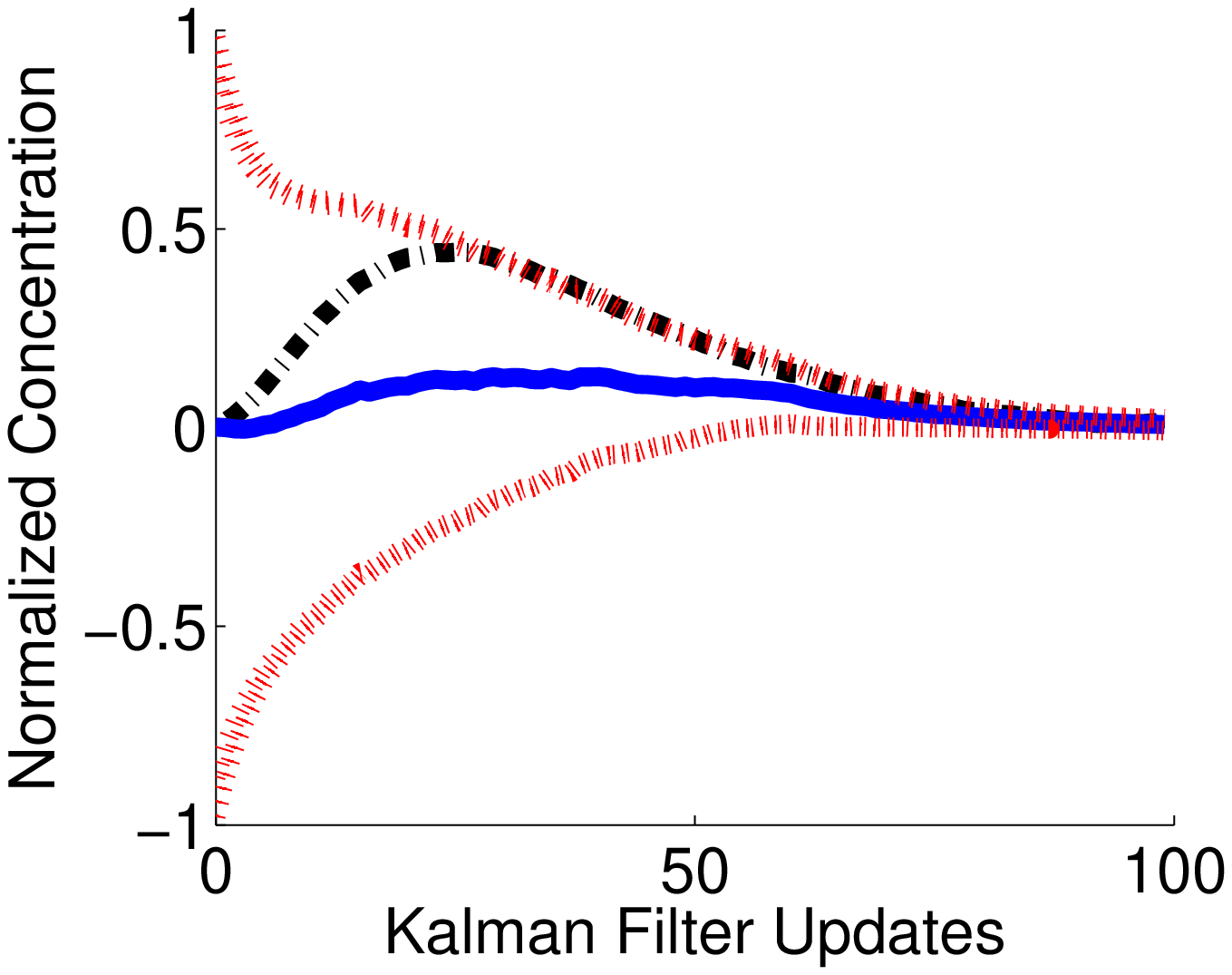}}
    \subfigure[Five Clusters]{\includegraphics[width=0.52\hsize]{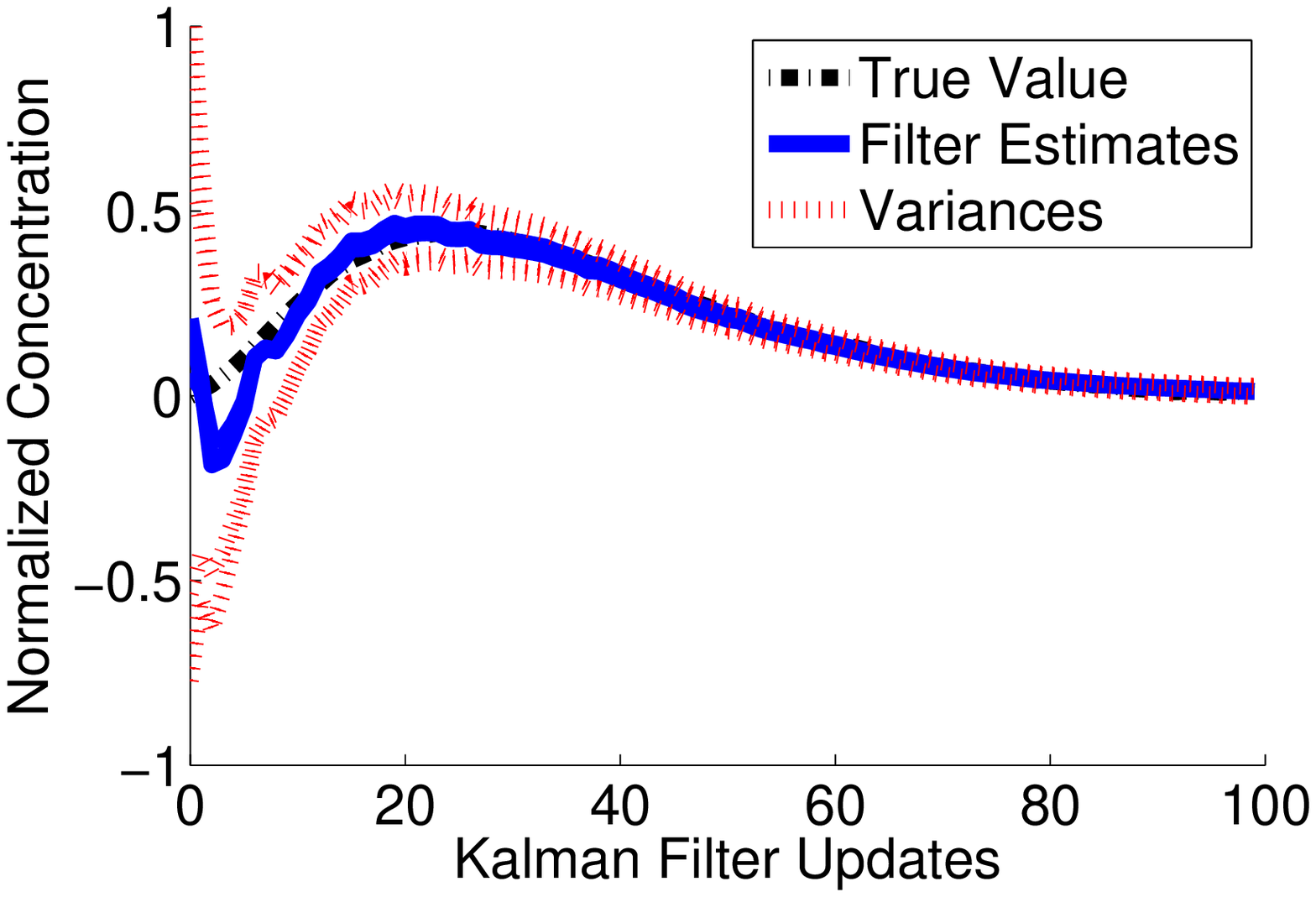}}
    \caption{Concentration estimates for room 49. Concentration estimates (solid line) are compared to the true value (dash-dot line). The dashed lines give the $\pm 3\sigma$ curves around the estimate. As it is clear from the plots, the strategy in
    which the consensus step is run using five clusters (b) is much better than using two clusters (a).}\label{fig:all_clusters_sim}
\end{figure*}

Numerical results are obtained by
running the Kalman filter interleaved with the consensus step,
see~\cite{JHK08}. We fix 10 iterations for the consensus step in
each cluster. Fig.~\ref{fig:all_clusters_sim} shows the
estimation result for 100 updates of the Kalman
filter\footnote{We assume that the consensus step is fast
compared to the contaminant spreading so that no compensation of
delay is required at the nodes while running the Kalman filter.
It is clear that for estimation, shortening the consensus step
is crucial in order to have a consistent estimate.} for both
clustering strategies described previously. In particular,
Figs.~\ref{fig:all_clusters_sim}a
and~\ref{fig:all_clusters_sim}b show the estimate (solid line)
of the concentration (the true value is shown with dash-dot
line) in room~49 made by all the sensors in the building. It can
be clearly seen that the estimate in
Fig.~\ref{fig:all_clusters_sim}a is not as accurate as the one
in Fig.~\ref{fig:all_clusters_sim}b. The reason is that the
consensus step for the case of four clusters on the second floor
converges much faster to the true covariance compared to the
case of two clusters. In comparison, if consensus is run for the
case of two clusters, it requires more than 500 iterations in
each consensus step to converge to the accuracy of
Fig.~\ref{fig:all_clusters_sim}b.

In the $5$ cluster case, all the nodes in the building have
accurate estimates of the contaminant concentration for rooms
located on the first floor. This is because sensors on the first
floor are strongly connected to one another and 10 iterations
are enough to converge to the true covariance (with only slight
corruption by the ``unconverged'' averaging on the second
floor).

These simulations show that the wave equation based clustering
provides an efficient distributed bottom-up methodology for
partitioning sensor networks and accelerating distributed
estimation algorithms.

%

\section{Mobile Sensor Networks}
\label{sec:search}

\begin{figure}[t!]
    \centering
 \includegraphics[width=0.95\hsize]{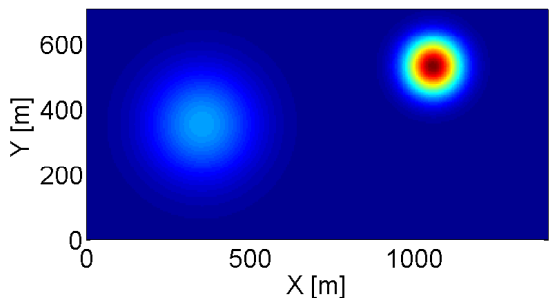}\\
    \caption{Prior belief map (distribution) for targets.}\label{fig:priormap}
\end{figure}

We demonstrate the utility of distributed partitioning for
computing the trajectories of mobile sensors/vehicles for the
purpose of efficiently searching a large area. In~\cite{George}
the authors have developed an algorithm to optimally search a
region, given a prior distribution that models the likelihood of
finding the target in any given location (see for example
Fig.~\ref{fig:priormap}). The trajectories are computed using a
set of ordinary differential equations given by,
\begin{equation}
    \dot x_{j}(t) = u_{j}(t)\,. \label{eq:dynveh}
\end{equation}
The above equation describes the dynamics of the $j$-th vehicle, where $x_{j}(t)$ and $u_{j}(t)$ are the position and the control input of the $j$-th vehicle at time~$t$ respectively. The authors prove that the control law
\begin{equation}
    u_{j}(t) = -u_{max}\frac{B_{j}(t)}{||B_{j}(t)||}\,,
    \label{eq:controlveh}
\end{equation}
efficiently samples the prior distribution for search. Here,
\begin{equation}
B_{j}(t) = \sum_{k}\frac{\Lambda_{k}S_{k}(t)\nabla
f_{k}(x_{j}(t))}{\langle f_{k},f_{k}\rangle}\,,
\label{eq:Beq}
\end{equation}
where~$f_{k}$ are the Fourier basis functions that satisfy the
Neumann boundary conditions on the domain to be searched and~$k$
is the corresponding basis vector number. The
quantities~$S_{k}(t)$ are governed by the following differential
equation,
\begin{equation}
    \frac{d S_{k}(t)}{dt} = \frac{\sum_{j=1}^{N}f_{k}(x_{j}(t))}{\langle f_{k},f_{k}\rangle} - N\mu_{k}\,, \label{eq:Sk}
\end{equation}
where~$N$ is the number of vehicles.

\begin{figure}[t!]
    \centering
\includegraphics[width=0.95\hsize]{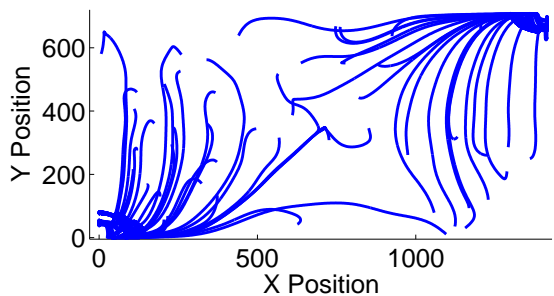}\\
    \caption{Trajectories generated using distributing spectral search algorithm that uses wave equation based clustering.}\label{fig:traj}
\end{figure}

In~\cite{George} the trajectories are computed a-priori for a
given distribution (belief map), using
Eqns~\ref{eq:dynveh},~\ref{eq:controlveh},~\ref{eq:Beq},~\ref{eq:Sk}.
Here we perform online computations for trajectories generation
in a distributed setting . The sum
$\sum_{j=1}^{N}f_{k}(x_{j}(t))$ over all vehicles in
Eq.~\ref{eq:Sk} is the centralized quantity that needs to be
computed in a distributed manner. At every time instant (every
time step of the Runge Kutta scheme), the vehicles are
partitioned into groups using the wave equation based clustering
algorithm and the sum in Eq.~\ref{eq:Sk} is computed over the
clusters and the solutions added. All the vehicles then compute
a piece of their trajectory for a predetermined horizon of time
(for a single Runge Kutta time step). These pieces of
trajectories for each agent are merged together to give
Fig.~\ref{fig:traj}. In this way, the mobile sensors group
themselves into clusters and compute their trajectories in a
distributed manner.

\section{Conclusions}
\label{sec:conclusions}

In this work, we have constructed a wave equation based
algorithm for computing the clusters in a graph. The algorithm
is completely distributed and at every node one can compute
cluster assignments based solely on local information. In
addition, this algorithm is orders of magnitude faster than
state-of-the-art distributed eigenvector computation algorithms.
Starting from a random initial condition at every node, one
evolves the wave equation and updates the state based solely on
the scalar states of neighbors. One then performs an FFT at each
node and computes the sign of the components of the eigenvectors
of the graph Laplacian. Complete eigenvector information can be
transmitted to each node using multi-hop communication. This
process is equivalent to spectral clustering.

The algorithm is also attractive from a distributed computing
point of view, where parallel simulations of large dynamical
systems~\cite{TuhinAdap} can be coupled to the distributed
clustering approach presented here, to provide scalable
solutions for problems that are computationally and
theoretically intractable. This application is the subject of
current research.

Wave equation based clustering is demonstrated on community
detection examples. Applications to multi-scale distributed
estimation and distributed search are also demonstrated.

Current work includes the extension of the wave equation based
algorithm for dynamic networks. This is clearly very important
in situations where the weights on the edges of the graph are
time varying. Examples of systems where dynamic graphs arise
include UAV swarms, nonlinear dynamical systems and evolving
social graphs.

\section*{Acknowledgements}
The authors thank Jos{\'e} Miguel Pasini, Alessandro Pinto and
Pablo Parrilo for valuable discussions and suggestions. The
authors also thank Andrea Lancichinetti for providing the
Fortunato benchmark graph. The authors would also like to thank
the anonymous reviewers for the constructive comments that
helped improve the paper.

\bibliographystyle{IEEEtran}
\bibliography{WaveEqPaper_Eigen_v4}
\end{document}